\documentclass[conference]{IEEEtran}
\usepackage{graphicx}
\usepackage{tabularx}
\usepackage{float}
\usepackage{amsthm}

\usepackage[centerlast,small,bf]{caption}
\usepackage{amssymb}
\usepackage{resizegather}
\usepackage{color}
\usepackage{subcaption}
\usepackage{color}
\usepackage{array}
\newcolumntype{M}[1]{>{\centering\arraybackslash}m{#1}}
\newcolumntype{P}[1]{>{\centering\arraybackslash}p{#1}}
\graphicspath{{folder/}{Fig/}}
\usepackage{multirow}
\usepackage{epstopdf}
\def\BibTeX{{\rm B\kern-.05em{\sc i\kern-.025em b}\kern-.08em T\kern-.1667em\lower.7ex\hbox{E}\kern-.125emX}}
\usepackage{array}
\usepackage{placeins}
\usepackage{afterpage}
\usepackage{caption}
\usepackage[papersize={8.5in,11in}, left=0.625in, right=0.625in, top=0.75in, bottom=1in]{geometry}
\usepackage[noadjust]{cite}
\usepackage{cleveref}

\begin{document}
\title{  
	Neuromorphic AI Empowered Root Cause Analysis of Faults in Emerging Networks 
}
\vspace{-4cm}
\author{
	
	{Shruti Bothe, Usama Masood, Hasan Farooq and Ali Imran}\\
	AI4Networks Research Center, Dept.\ of Electrical \& Computer Engineering, University of Oklahoma, USA\\
	Email: \{shruti, usama.masood, hasan.farooq, ali.imran\}@ou.edu\\
	\fontsize{10}{10}\selectfont\rmfamily\itshape
	\vspace{-0.5cm}
}

\maketitle

\begin{abstract}
Mobile cellular network operators spend nearly a
quarter of their revenue on network maintenance and management. A significant portion of that budget is spent on resolving
faults diagnosed in the system that disrupt or degrade cellular services. Historically, the operations to detect, diagnose
and resolve issues were carried out by human experts. However,
with diversifying cell types, increased complexity and growing cell
density, this methodology is becoming less viable, both technically
and financially. To cope with this problem, in recent years, research on self-healing solutions has gained significant momentum.
One of the most desirable features of the self-healing paradigm
is automated fault diagnosis. While several fault detection and
diagnosis machine learning models have been proposed recently,
these schemes have one common tenancy of relying on human
expert contribution for fault diagnosis and prediction in one
way or another. In this paper, we propose an AI-based fault
diagnosis solution that offers a key step towards a completely
automated self-healing system without requiring human expert
input. The proposed solution leverages Random Forests classifier,
Convolutional Neural Network and neuromorphic based deep
learning model which uses RSRP map images of faults generated.
We compare the performance of the proposed solution against
state-of-the-art solution in literature that mostly use Naive Bayes
models, while considering seven different fault types. Results show
that neuromorphic computing model achieves high classification
accuracy as compared to the other models even with relatively
small training data.

\end{abstract}


\vspace{-0.24cm}
\section{Introduction}
\vspace{-0.17cm}

Unprecedented growth in the cellular network usage in the
last decade has made system operations more complex. Billions
of dollars are spent in the United States alone by cellular
operators to detect and diagnose faults \cite{Ahmad_Selfhealing}. With the genesis of
5G ultra-dense network, fault management will be a primary
challenge. It is comprehensible to foresee that with such an
amalgam of technologies and operational complexity required
to configure, operate, optimize and maintain the networks the
biggest challenges in the emerging cellular networks \cite{BSON_vision}  would
be to automate the fault diagnosis process.

Even single fault in the network system always produces
a large amount of alarm information. Operators then have to
depend on domain experts to diagnose the exact cause and
devise a solution. The drawback of this is that it is very time
consuming and is prone to human errors given the unfathomable complexity of the system. From the perspective of a
network operator, achieving timely and accurate diagnosis of
the cause of the faults is critical for both improving subscriber perceived experience and maintaining network reliability.

Network applications necessitate that network fault issues be
resolved in a very short time. The complexity of fault diagnosis
has led to emergence of current solutions that use a combination of expert input and an automated system. Although
research is on-going, most of these fault diagnosis systems
were designed on make-shift and incoherent basis. This involved simply dissemination of human expert knowledge into a
system that has the capability of solving problems in a specific
domain. But, by taking into consideration future emerging cellular networks, their complexity and shrinking profit margins;
minimizing human interaction for fault diagnosis for Self Organizing Networks (SON) would unquestionably be desirable \cite{Ahmad_Selfhealing,BSON_vision,masood2018deep}. This means that automation for reducing costs, handling
complexity and maximizing resources efficiency will not only
become a necessity, but future cellular networks will depend
on it \cite{Ahmad_Selfhealing}. The main task of an automated fault diagnosis tool
is to identify the root cause of problem. This paper focuses on
the fault diagnosis process since the accuracy and timeliness
of the fault management process is highly dependent on it. 

\begin{figure*}[t!]
	\centering
	\includegraphics[trim={0cm 6.9cm 0cm 0cm},clip,width=5.7in,height=2in]{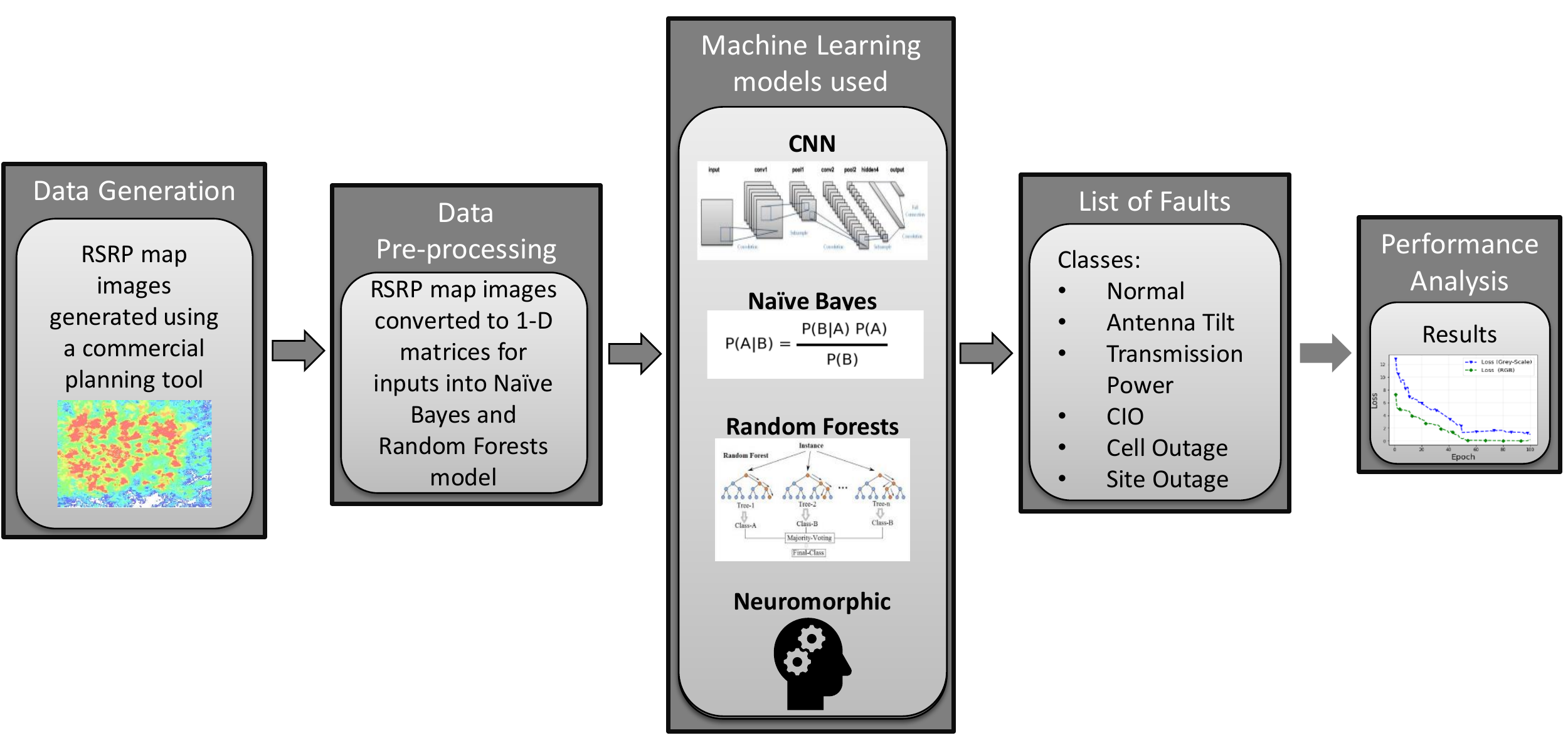}
	\caption{Proposed framework for fault diagnosis in cellular networks}
	\label{f}
\end{figure*}

\subsection{Relevant Work}
Several recent studies have explored fault diagnosis in cellular networks and primarily focus on certain predefined faults.
Authors in \cite{BN} have proposed the use of Bayesian networks to
diagnose faults. In \cite{Data_mining}, authors described the functionalities that
targeted automated troubleshooting in Radio Access Network
(RAN) by using ‘if-then’ rules. Similarly, in \cite{Auto_diag}  an approach for
automated diagnosis in Universal Mobile Telecommunications
System (UMTS) networks using a Bayesian network (BN) is
proposed. Alarms are modeled as discrete random variables
with two states: OFF or ON. KPIs are modeled as discrete
random variables with two, three, or more states. The authors
in \cite{per_indi} propose a system for automated diagnosis using Naive
Bayesian (NB) classifier, which is used to identify the cause
of fault in 3G, GSM/GPRS or multi-systems networks. In \cite{Self_healing}  authors presented a self-healing framework for detecting
and compensating cell outages and evaluated it in realistic
environment. Authors in \cite{Learning_model}  have compared thresholds obtained
with 3 different algorithms such as knowledge-based method
(TEXP), Entropy Minimization Discretization (EMD) and Selective Entropy Minimization Discretization. In \cite{OSS},  authors
used unsupervised clustering for detection and diagnosis of
the root cause of faults. Five classification algorithms, namely
Chi-squared automatic interaction detection (CHAID), quick
unbiased efficient statistical tree (QUEST), Bayesian network, support vector machine (SVM) and classification and regression tree (CRT) were used. Authors in \cite{integrate} and \cite{Auto} have designed a framework for detection and diagnosis to find the root cause which is based on expert knowledge. In \cite{13}, a few improvements to this framework are discussed. The work has included more sophisticated profiling and detection capabilities.
\subsection{Contributions}
Although all the above mentioned works contribute on similar lines in terms of automating the fault diagnosis process, the novel contributions of this paper are outlined below:
\begin{enumerate}
	\item To the best of the authors' knowledge, none of the existing studies have used Convolutional Neural Networks (CNN), Random Forests (RF) and neuromorphic computing models for fault diagnosis.
	\item Existing studies use small pre-existing data sets limiting
	the number and types of faults that can be diagnosed.
	To overcome this limitation, in this study for the first
	time we have generated a synthetic data set. The entire
	process of generating the dataset has been automated in a ray-tracing based commercial planning tool.
	\item Most existing studies still rely on input/domain knowledge from human expert in one form or other. In
	the proposed solution, no input from human experts is
	needed. The models are purely trained on Reference
	Signal Received Power (RSRP) map images, which are
	labeled as faults. This paper presents a pure deep learning
	model where no additional external information provided
	by experts is needed.
	\item The final contribution of the paper is a comparison
	between NB solution that is the most widely investigated approach in literature for fault diagnosis, and our
	proposed CNN, RF and neuromorphic based solutions.
	Results show that all the proposed models outperform
	the prevailing NB fault diagnosis solutions and RF even
	outperforms CNN. 
\end{enumerate}
	
The rest of the paper is structured as follows: Data generation
and pre-processing methodology is presented in Section II.
Section III presents the CNN, NB, RF and neuromorphic model frameworks used in this paper. Numerical results are presented in Section IV. Section V concludes the paper.

\section{Data Generation and Pre-Processing}

\subsubsection{Data Generation}
\begin{figure}[h!]
	\vspace{-0.1cm}
	\centering
	\includegraphics[width=1.6in,height=1.6in]{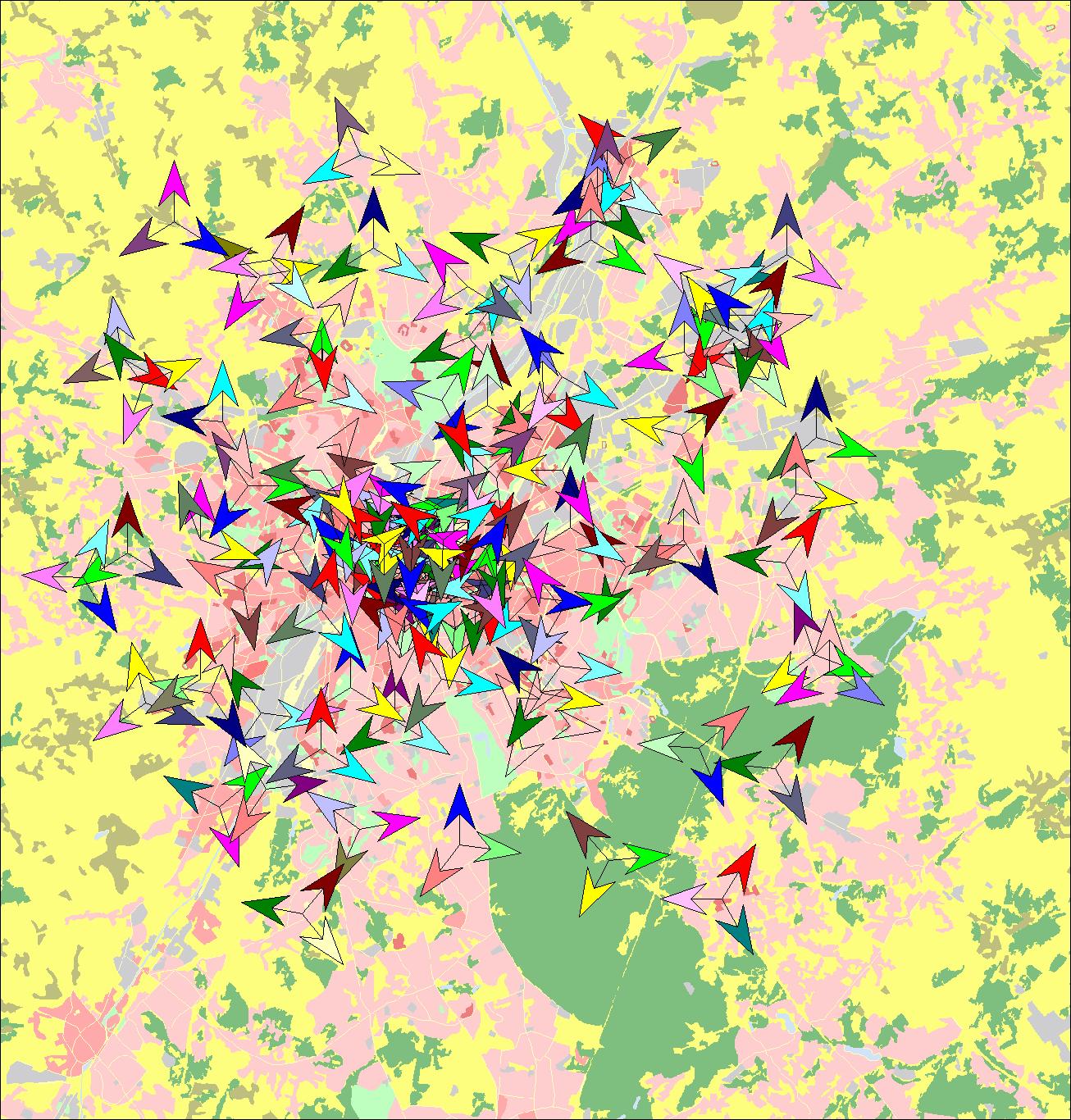}
	\caption{Spatial representation of the Base-Station positions}
	\label{clutter}
	\vspace{-0.1cm}
\end{figure}

The dataset used in this paper consists
of RSRP map images generated using a ray tracing based commercial planning tool used for wireless network design and optimization.
The area under consideration is the city of Brussels spanning over 800 $km^2$ . As seen in \Cref{clutter}, 291 transmitters have been positioned around the city over 15 clutter classes, which have been included in this framework
and each base-station (BS) has 3 transmitters connected to it.
For our analysis, we have considered 120 transmitters i.e. 40
BS. Also, to minimize the boundary effect, all transmitters are
kept functional.

\begin{figure*}[t!]
	\centering
	\begin{subfigure}{0.24\textwidth}
		\includegraphics[width=\textwidth,height=3.7cm]{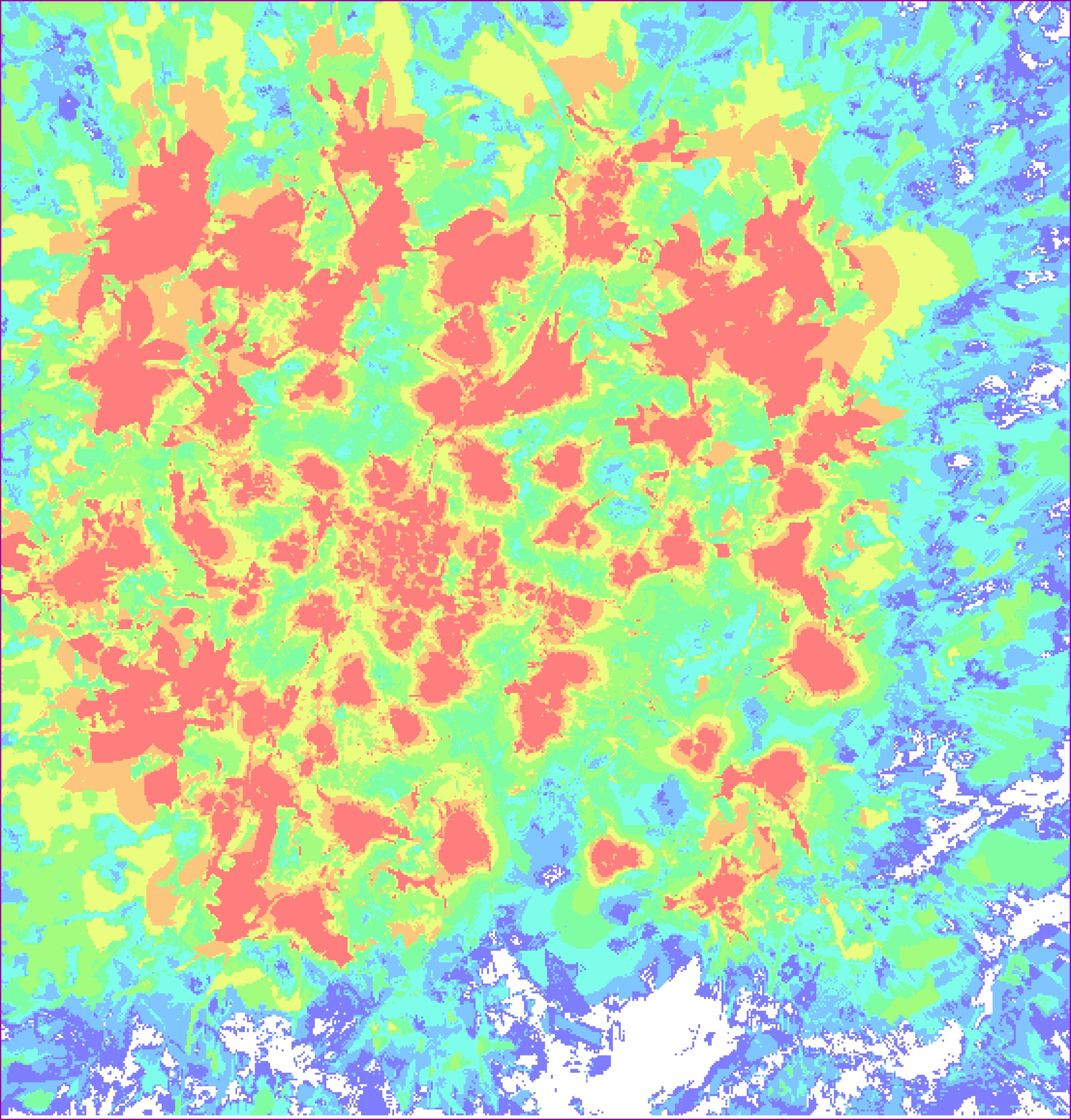}
		\caption{Normal Scenario}
	\end{subfigure}
	\begin{subfigure}{0.24\textwidth}
		\includegraphics[width=\textwidth,height=3.7cm]{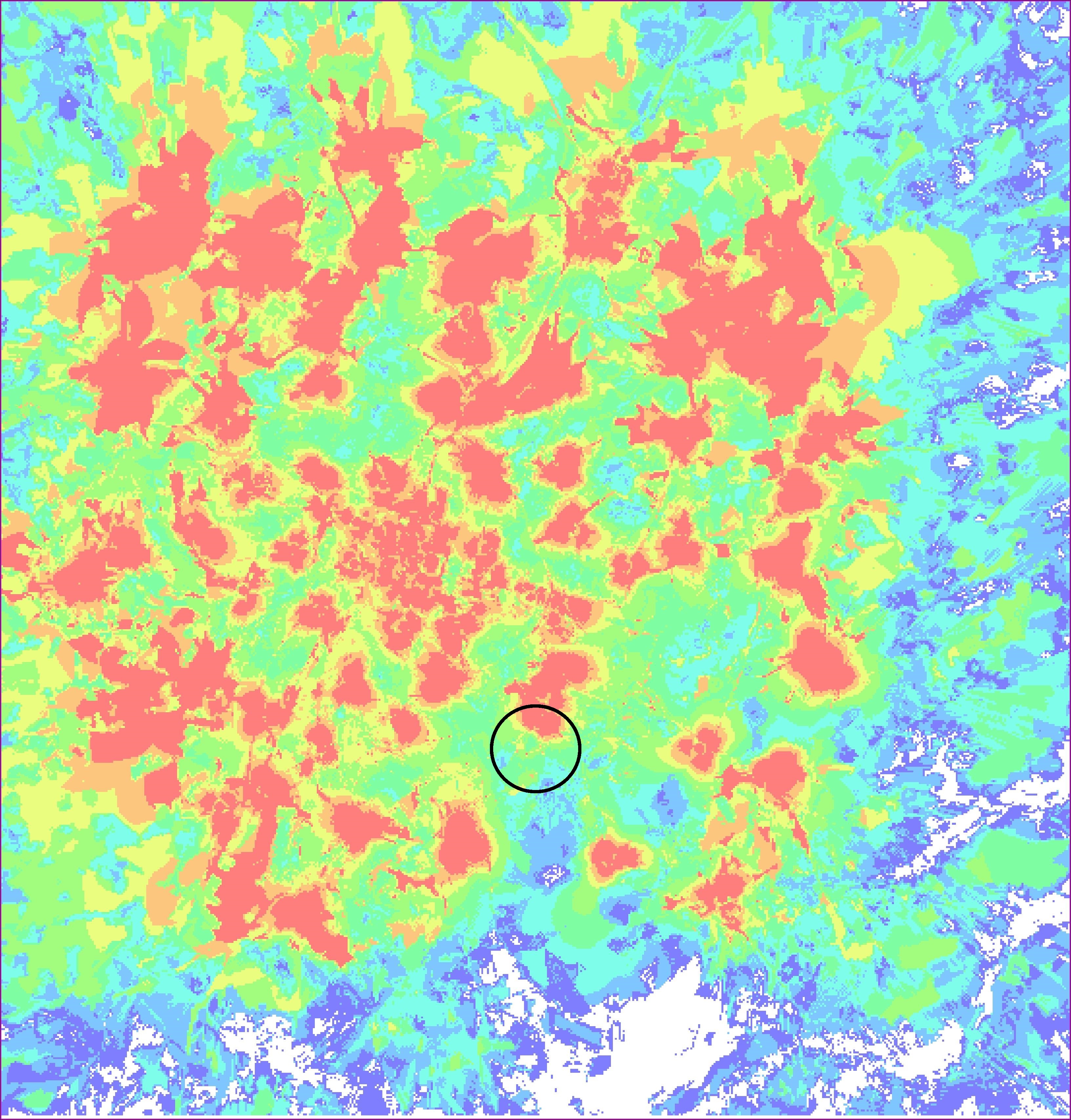}
		\caption{Transmitter Off}
	\end{subfigure}
	\begin{subfigure}{0.24\textwidth}
		\includegraphics[width=\textwidth,height=3.7cm]{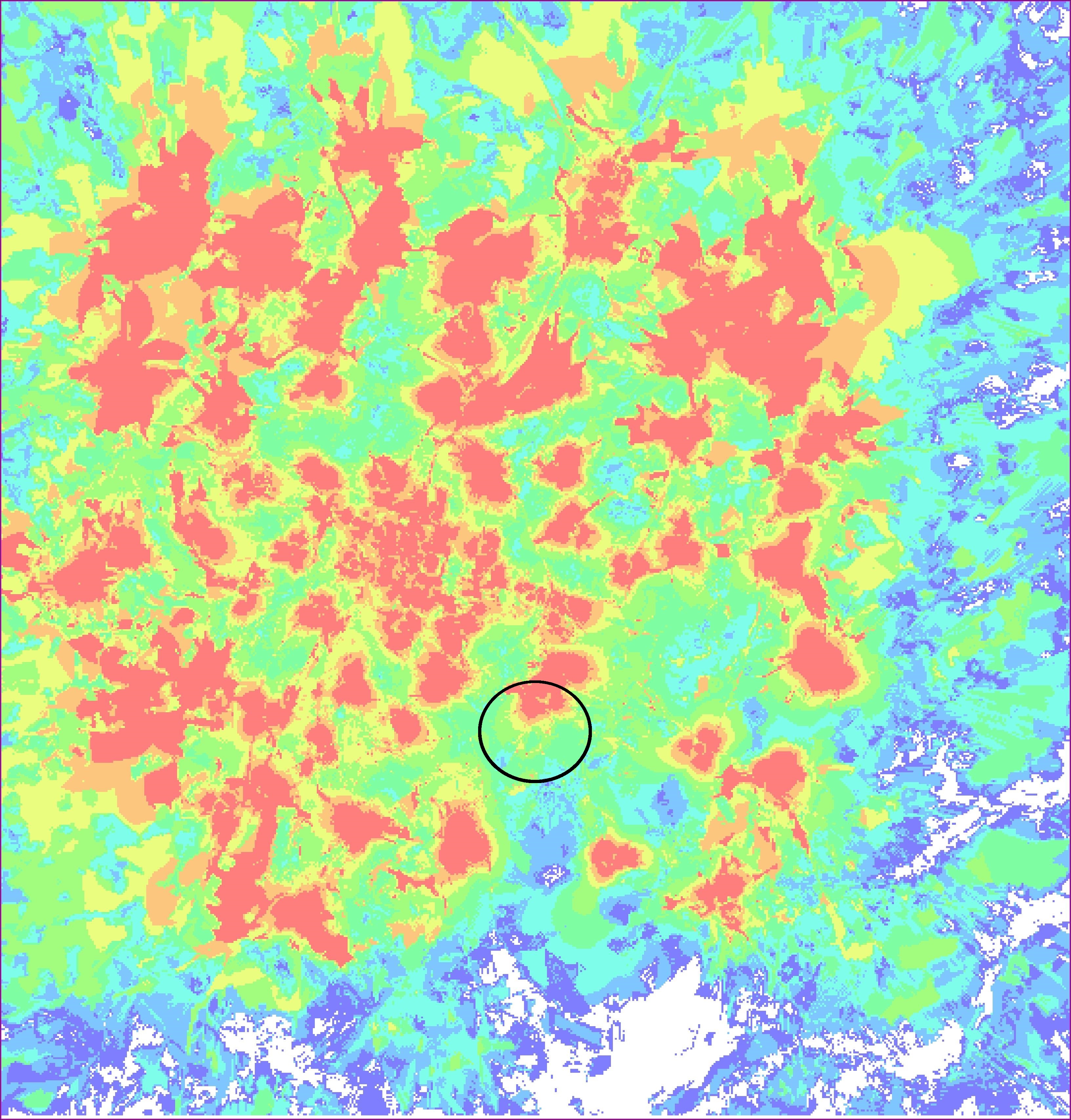}
		\caption{Site Outage}
	\end{subfigure}
    \begin{subfigure}{0.24\textwidth}
    	\includegraphics[trim={0cm 11.59cm 27cm 0.03cm},clip,width=\textwidth,height=4.2cm]{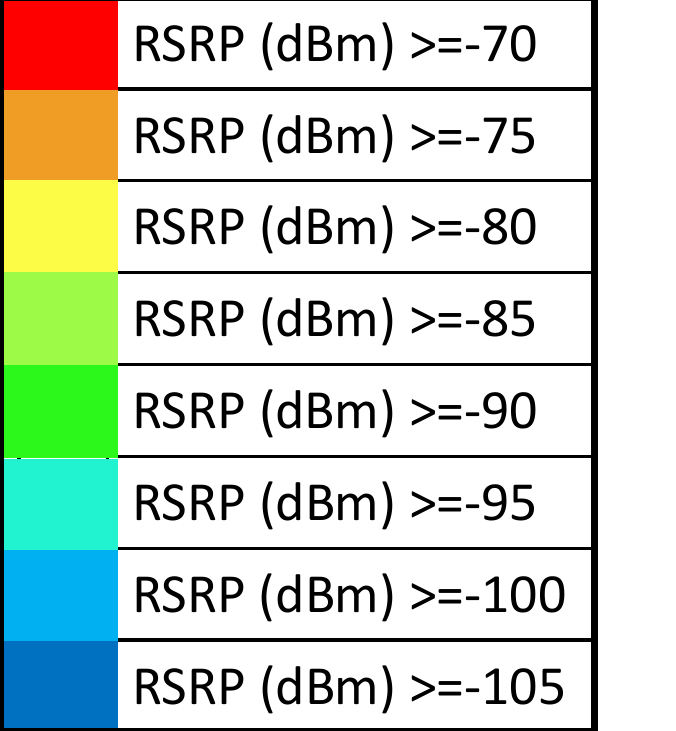}
    	
    \end{subfigure}
	\caption{Comparision of different faults generated in the network}
	\label{compare}
	\end{figure*}

Table I summarizes the network default settings before any
fault is induced. In this data generation technique, 7 faults have
been generated one at a time. These are generated in the BS in
the area of interest. RSRP map images were created for each
instance when an error is triggered and are later fed into the
aforementioned models. \Cref{compare} represents the RSRP map images
for ‘transmitter off’ and ‘site outage’ faults in one of the BS
as encircled.

Seven most commonly occurring faults in real networks were simulated to generate the synthetic data as described below:
\begin{enumerate}
\item  { \bf Cell Outage} (C.O): This error was created by turning each transmitter off one at a time. 
\item {\bf Site Outage} (S.O): A BS consisting of 3 transmitters was turned off at one time. 
\item {\bf Transmission Power} (TxP): An error in the transmission power of the BS was induced. The default value of 43 dBm was varied between 25 dBm to 35 dBm with a step size of one to trigger the error.
\item {\bf CIO Positive} (CIO+): To account for handover parameter error, the Cell INdividual Offset (CIO) was increased to 10 dB from its default value of 0 dB.
\item {\bf CIO Negative} (CIO-): A second error in CIO was created by reducing the CIO value from its default value of 0 dB to negative 10 dB.
\item {\bf Antenna Uptilt} (A$_U$): The tilt angle of the antenna was increased from 0 degrees to 25 degrees.
\item {\bf Antenna Downtilt }(A$_D$): Antenna tilt was decreased from its original value of 0 degrees to negative 25 degrees.
\end{enumerate}

\subsubsection{Data Pre-Processing}
\noindent The dataset consists of 1961 colored images of size 553x578 split into 8 classes namely, Normal Scenario, C.O., S.O., TxP, CIO+, CIO-, A$_U$ and A$_D$. The train test division is 70:30. The advantage of using RSRP
map images over raw data for fault diagnosis, as most prior
studies do, is that there is no need for identification of smart
input features or labeling required from experts. No preprocessing of the data is done before feeding it to the CNN.
Since images are simply RSRP maps, in real network such
maps can be generated using Minimization of Drive Test (MDT) data and then fed to the
proposed diagnosis framework.

\begin{table}[!h]
	\caption{Network Scenario Default Settings}  
	\renewcommand\arraystretch{1.4}
	\centering
	\begin{tabular}{|m{2cm}|m{1.5cm}|m{2.5cm}|}
		\hline
		\multicolumn{2}{|c|}{ \bfseries\sffamily System Parameters}
		& \multicolumn{1}{c|} { \bfseries\sffamily Values} \\ \hline
		\multicolumn{2}{|c|}{Cellular Layout} & \multicolumn{1}{c|}{ 120 Macrocell sites} \\ \hline
		\multicolumn{2}{|c|}{Sectors} & \multicolumn{1}{c|}{ 3 sectors per BS} \\ \hline
		\multicolumn{2}{|c|}{Simulation Area  } &\multicolumn{1}{c|} { 800 km$^2$} \\ \hline
		
		\multicolumn{2}{|c|}{Path Loss Model } & \multicolumn{1}{c|}{ Ray-Tracing} \\ \hline
		\multicolumn{2}{|c|}{Land Cover (Clutter) Types } & \multicolumn{1}{c|}{ 15 different classes} \\ \hline
		\multicolumn{2}{|c|}{BS Transmit Power  } & \multicolumn{1}{c|}{ 43 dBm} \\ \hline
		\multicolumn{2}{|c|}{Cell Individual Offset  } & \multicolumn{1}{c|}{  0 dBm} \\ \hline
		\multicolumn{2}{|c|}{Antenna Tilt   } & \multicolumn{1}{c|}{ 0 deg} \\ \hline
		\multicolumn{2}{|c|}{Antenna Gain   } & \multicolumn{1}{c|}{ 18.3 dBi} \\ \hline
		\multicolumn{2}{|c|}{Carrier Frequency  } & \multicolumn{1}{c|}{ 2100 MHz} \\ \hline
		
		\multicolumn{2}{|c|}{Geographic Information  } & { Ground Heights +  Building Heights  +  Land Use Map } \\ \hline

	\end{tabular}
\end{table}

\begin{figure}[h!]
	\centering
	\begin{subfigure}{0.224\textwidth}
		\includegraphics[width=\textwidth,height=3.7cm,height=3.7cm]{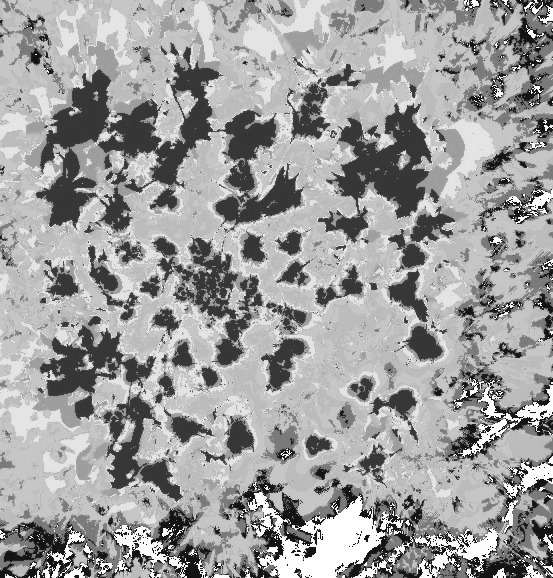}
		
	\end{subfigure}
	\begin{subfigure}{0.124\textwidth}
		\includegraphics[trim={0cm 11.5cm 27cm 0cm},clip,width=\textwidth,height=3.7cm,height=3.7cm]{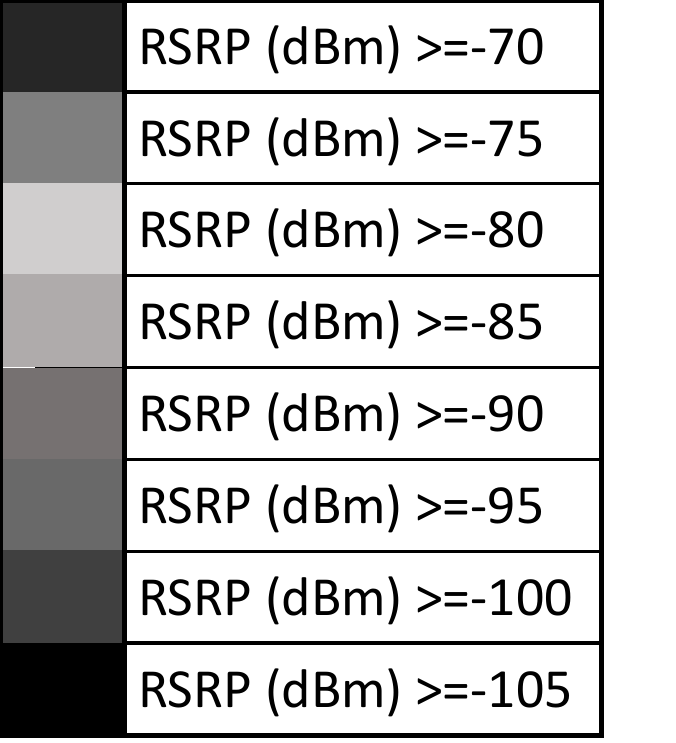}
		
	\end{subfigure}		
	
	\caption{Grayscale RSRP map used as input to CNN model}
	\label{figgrey}
\vspace{-0.6cm}
\end{figure}

However, since NB and RF do not take images as inputs, 1-D
matrices corresponding to the RSRP map images are generated.
The associated label for each class are induced in the matrix.
A digital grayscale image is represented by a pixel matrix. In
such an image each pixel is represented by one integer from \{0,...,255\}. Zero represents black pixels and 255 as white. RGB images are represented with three grayscale images
matrices (one for each red, green and blue color). \Cref{f} summarizes the framework for fault diagnosis model. All the RSRP map images are also converted to grayscale to be fed in these models to compare the networks' performance between grayscale and RGB input data. A representation of this is shown in Fig \ref{figgrey}.

	
\section{Proposed Model Framework}

\subsection{Convolutional Neural Network Model}
CNNs are Artificial Neural Networks (ANN) that have
been most commonly used for analyzing images. This pattern
detection is what makes CNNs so useful for images analysis.
The hidden convolutional and non-convolutional layers makes
them different from the standard multi-layer perceptron(MLP).
Since its introduction, CNNs have been applied in various research such as sentence classification \cite{sent_class}, face recognition \cite{face}, traffic signal recognition \cite{traffic} and many more. 3 operations
mainly:convolution, transformation and pooling are performed on the input image.
In this model, we have provided RGB and grayscale RSRP map images.

The pre-defined number of filters in the convolutional layers
learn the features of the input image. During a convolution,
the filters effectively slide over the input feature map grid
horizontally and vertically, one pixel at a time, extracting each
corresponding tile. For each filter-tile pair, CNN performs
element-wise multiplication of the filter matrix and the tile
matrix, and then sums all the elements of the resulting matrix
to get a single value. Each of these resulting values are the
outputs in the convoluted feature matrix. During training, the
CNN 'learns' the optimal values for the filter matrices that
enable it to extract meaningful features from the input feature
map. As the number of filters applied to the input increases,
so does the number of features the CNN can extract. However,
the trade-off is that filters compose the majority of resources
expended by the CNN, so training time also increases as more
filters are added.

Following each convolution operation, the CNN applies a
Rectified Linear Unit (ReLU) transformation to the convoluted
feature, in order to introduce nonlinearity into the model. The
equation for ReLU can be written as follows:
\vspace{-0.1cm}
\begin{equation}
\bf f(x) = max(0,x)
\vspace{-0.4cm}
\end{equation} 

where x is the input to the neuron. ReLU can allow the
model to account for non-linearities and interactions.

Neurons are only locally connected by filters, followed by a pooling layer of fixed size 2x2. For this model we have used max-pooling. This max-pooling layer down-samples the
convoluted feature, reducing the number of dimensions of the feature map, while still preserving the most critical feature information and to control over fitting. In a fully connected layer all the neurons consider every activation in the previous layer. Each layer learns its weights and biases using gradient descent in small mini-batches of training samples. In this model, the batch size to train is 32 and the number of output classes is 8. Learning rate is set to 0.001 and the maximum number of epochs is 100.

Since the neurons in a fully connected layer have connections to all activations in the previous layer, their activations
can be computed with a matrix multiplication followed by a
bias offset. In the convolutional layer the neurons are connected
only to a local region in the input, and only those many the
neurons in a convolutional volume share parameters. However,
the neurons in both layers still compute dot products, so their
functional form is identical. The last fully-connected layer
holds the output.

\subsection{Naive Bayes Classifier Model}

NB classifiers are a collection of classification algorithms
with strong (naive) independence assumptions between the
features in which every pair of feature being classified is
independent of each other. NB is particularly useful for large
data sets. To input the data-set consisting of images, a matrix
in which each row is a 1-D array associated to an image is
created. Next, an array which contains the associated label for
each class, i.e all the fault labels used in this paper as described
above was added to the matrix. The NB classifier combines
this model with a function which maps an observation to an
appropriate action. This classifier is the function that allocates
the class label. The classifier used in this paper is Gaussian
NB classifier and can be expressed as: 
\vspace{-0.1cm}
\begin{equation}
P(x_i \mid y) = \frac{1}{\sqrt{2\pi\sigma^2_y}} \exp\left(-\frac{(x_i - \mu_y)^2}{2\sigma^2_y}\right)
\end{equation}
\noindent where $x_i$ denotes the feature vector and y is the class variable. $\sigma_y$ and $\mu_y$ are estimated using maximum likelihood. The class with the highest probability is considered as the most likely class. This is also known as Maximum A Posteriori (MAP). The MAP for a hypothesis is:
\begin{align}
MAP(H) &=max( P(H|E) )\\
      & =  max( (P(E|H)*P(H))/P(E))\\
       &= max(P(E|H)*P(H))
\end{align}
P(E) is evidence probability, and it normalizes the result.

The advantage of using NB is that it performs well in multiclass prediction and is fast to predict class of test data set.
In cases where assumption of independence is valid, a NB
classifier performs better compared to logistic regression and
less training data is needed. But the disadvantage of using NB
is that if categorical variables have a category in test data set,
which were not observed in training data set, then the model will
assign a 0 (zero) probability and will be unable to make a
prediction. This is often known as “Zero Frequency”.

\subsection{Random Forests Classification}

A third model in the proposed framework is the RF classifier.
A Decision Tree (DT) builds a set of rules that predict a class
for classifying complex situations. DT when used as ensemble
can even outperform neural networks in some cases \cite{RF}.  The
most popular ensemble technique is RF. A RF is a collection
of DT whose results are aggregated into one final result. RFs
limit over fitting without substantially increasing error due to
bias by training on different samples of the data. Boostrap
resampling method aids in extracting multiple samples from the
original samples and to construct sub-datasets. The sub dataset
are then used to form the base decision tree and train on it.
This method is a statistical technique for estimating quantities
about a population by averaging estimates from multiple small
data samples. A second method is by using a random subset
of features. This means, for a set of 8 features, RF will only
train on a certain number of features in each model. Thus,
in each tree a minimum number of random features can be
utilized. In a forest, eventually most or all of the features would have been included making them much more robust than a single decision tree. RF also offers a good feature selection indicator. RF use Gini importance or mean decrease in impurity (MDI) to calculate the importance of each feature. The Gini coefficient (GC) measures the inequality among values of a frequency distribution where zero expresses perfect equality and 1 or 100\% expresses maximal inequality among values. The RF classifier's parameters that have to be set, consist of number of trees, number of features, impurity function and stop criteria. Generally, the default values have been used for this. The maximum number of iteration chosen is 100. The depth of tree is limited  to 5 levels. The overall accuracy and Kappa Coefficient obtained by RGB images are 77.82\% and 0.63, respectively. For the grayscale images, these values are 87.16\%; and 0.79, respectively.

\subsection{Neuromorphic Computing model}

The goal of this approach is to assess the similarity of the
decision making mechanisms of convolutional neural networks
and corresponding mechanisms in the human cortex. This
tool greatly helps in understanding the dynamic method of
learning and development in the brain. This can be an inspiration which can be applied on generic cognitive computing.
In late 1980, this concept was developed by Carver Mead {\cite{Mead}},  describing the use of very-large-scale integration (VLSI)
systems containing electronic analog circuits to mimic neurobiological architectures present in the nervous system. We
may expect neuromorphic technologies to deliver a range of
applications more efficiently than conventional computers. For
instance, to deliver speech and image recognition capabilities
in smart phones. Currently, such capabilities are available only
using powerful cloud resources to implement the recognition
algorithms.
\subsubsection{Nengo}

Nengo \cite{Nengo}  is a neural simulator based on a
framework called the Neural Engineering Framework (NEF).
The Neural Engineering Framework (NEF) is a set of theoretical methods that are used in Nengo for constructing
neural models. The NEF is based on Eliasmith and Anderson’s
(2003) book from MIT Press. Nengo is highly extensible and
flexible. One can define their own neuron types and learning
rules, get input directly from hardware, build and run deep
neural networks. It is a large-scale modeling approach that can
leverage single neuron models to build neural networks with
demonstrable cognitive abilities.

\subsubsection{Nengo Neural Network}
Synapses are unidirectional connections that neurons use to communicate. When a neuron
spikes, the neurotransmitter released across the synapse, causes
some quantity of current to be transmitted in the postsynaptic
(downstream) neuron. Factors affecting the amplitude of the
current can be summarized in a scalar connection weight. To
compute any function, the weights of the connection are set
as the product of the decoding weights for that function in
the first population, the encoding weights for the downstream
population, and any linear transform.
Non-neural information (sensory inputs and motor outputs)
are represented by the Node. A complex experimental setup that provides input and responds to the output can be modeled. The Connection describes how ensembles and nodes are
connected. The Probe gathers data during a simulation for
later analysis. The Network encapsulates a functionally related
group of interconnected nodes and ensembles. The Model
encapsulates a Nengo model. Several signals are created for
each high-level Nengo object; i.e. the simulator creates signals
that represent the high-level input signal that will be encoded
to input currents, and the encoding weights. The ensemble also
contains a neural population, for which the simulator creates
signals that represent input currents, bias currents, membrane
voltages, and refractory times for each cell.

\section{Performance analysis}

\begin{figure}[h!]
	\centering
	\includegraphics[width=0.85\linewidth]{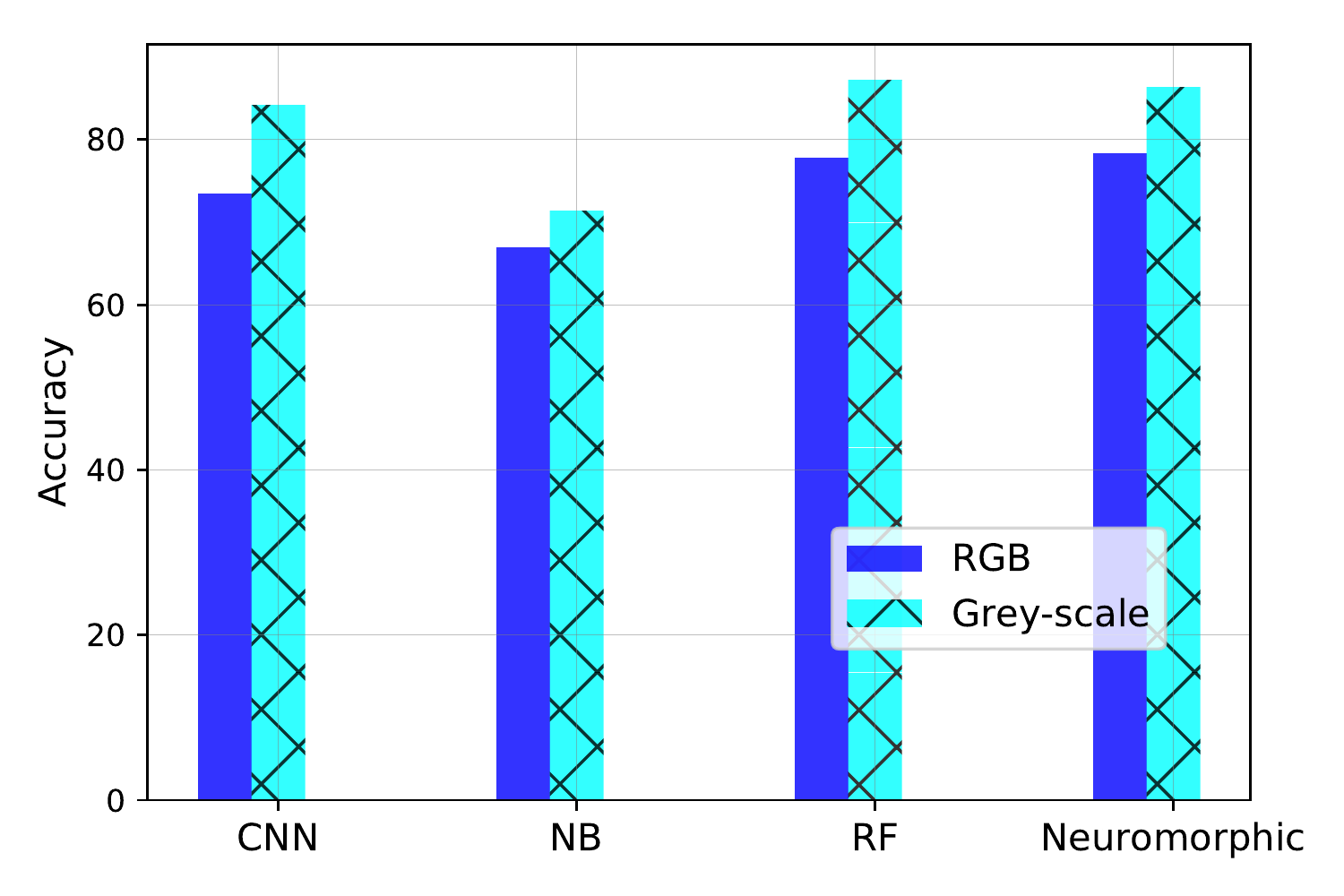}
	\caption{Graph of Accuracy for grayscale and RGB images}
	\label{accuracy}
	\vspace{-0.1cm}
\end{figure}

All the models explained earlier perform better for grayscale images than RGB images. As seen in Fig \ref{accuracy}, a visible
difference in accuracy can be noted. This difference in accuracy
can be attributed to the increase in the number of kernels. The
kernel size for grayscale image is k x k x 1 where as it
is k x k x 3 for an RGB image. Depending on the number
of kernels, the number of parameters increases proportionally.
Furthermore, large dimensional vectors are almost equally
spaced and hence are difficult to separate by a classifier. This
can be accredited to the curse of dimensionality phenomena.
Although no substantial information is added in an RGB image,
it in turn adds to more compute and memory intensive training
time due to large number of input parameters. Hence, feeding
these model with a grayscale image rather than an RGB image
results in a better classification accuracy.

\vspace{-0.3cm}

\begin{figure}[h!]
	\vspace{-0.35cm}
	\centering
	\includegraphics[width=2.25in,height=1.75in]{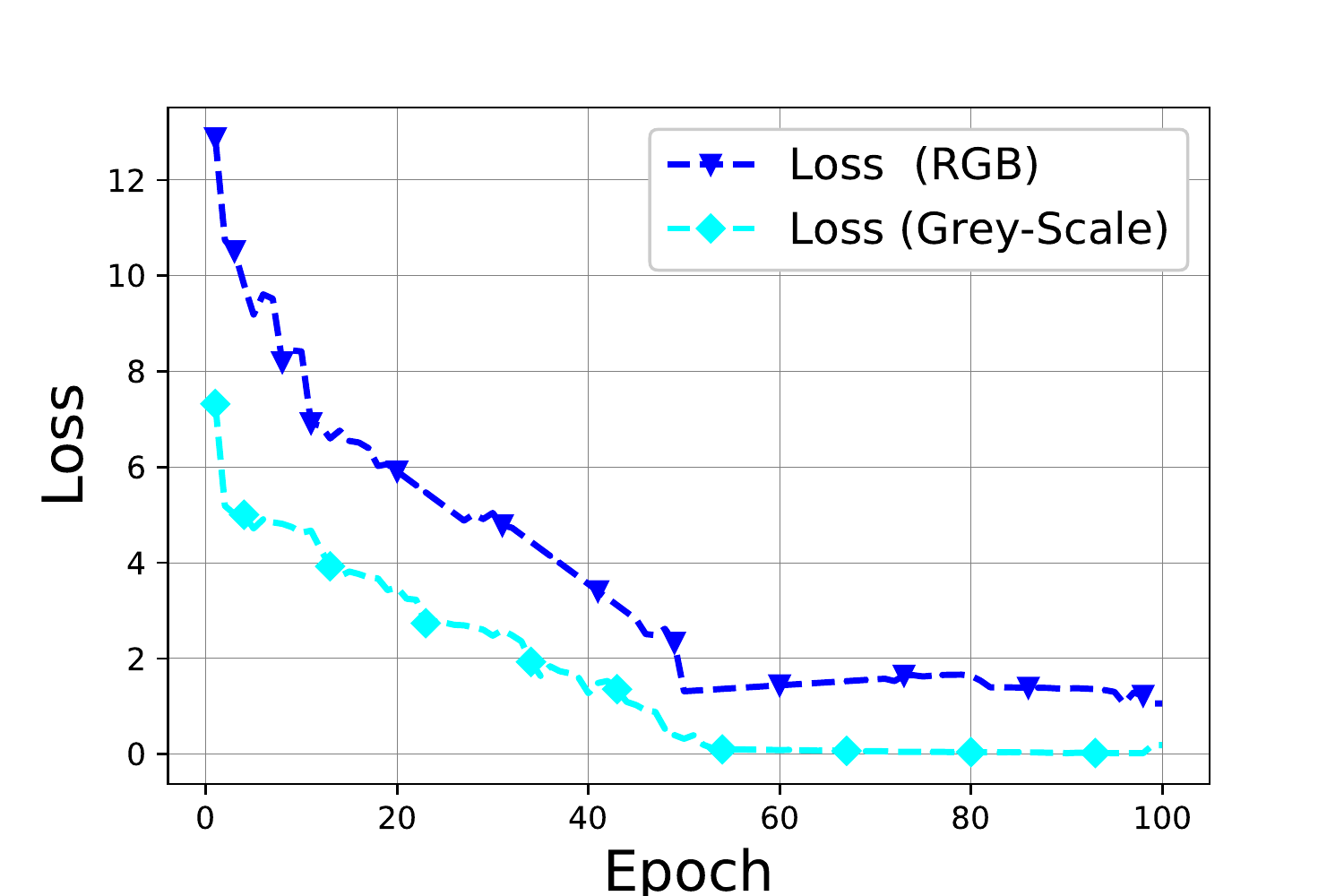}
	\caption{CNN model loss values for grayscale and RGB images}
	\label{loss}
	\vspace{-0.35cm}
\end{figure}

\begin{figure}[h!]
	\centering
	\includegraphics[width=2.7in,height=1.6in]{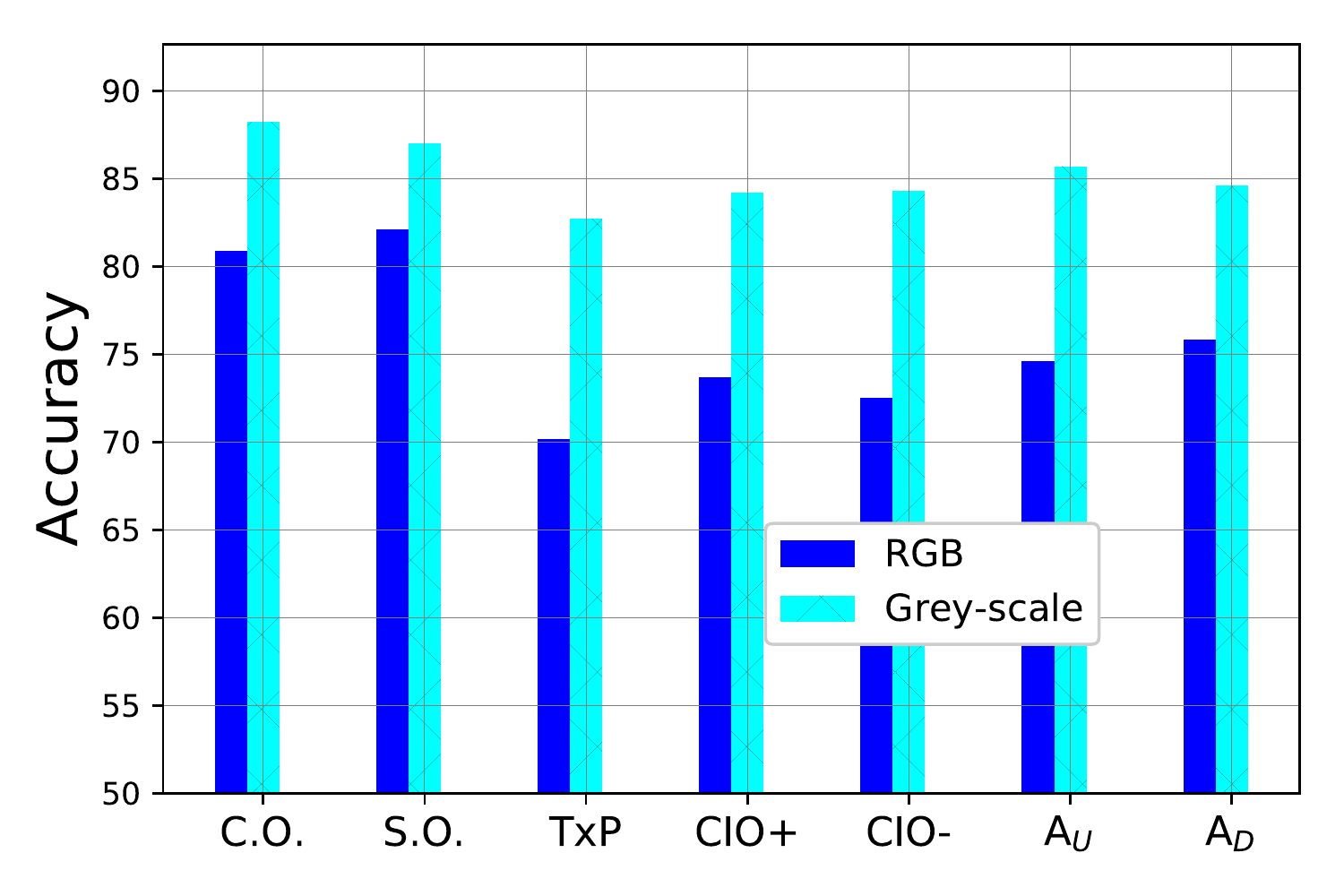}
	\caption{RF prediction accuracy for individual faults}
	\label{fault}
	\vspace{-0.25cm}
\end{figure}

Advantages of neuromorphic computing as compared to
the other approaches are execution speed, adaptability, energy
efficiency, the ability to learn, robustness against local failures
and diverse cell types at individual nodes. There are primarily
two ways in which the scene perception or image recognition is
performed in the human brain:
\begin{enumerate}
	\item an object-centered approach, in which components of a scene are segmented and serve
as scene descriptors;
	\item  space-centered approach, in which
spatial layout and global properties of the whole image or place
act as the scene descriptors.
\end{enumerate}
 The RF model has the highest
classification accuracy followed by the neuromorphic model,
CNN and NB. Some classifier combination techniques like
ensembling, bagging and boosting may improve NB classification accuracy but these methods would not help since their
purpose is to reduce variance. NB has no variance to minimize.
Accuracy of CNN saturates after a particular number of epoch as the model has limited number of data samples to train from. Neural Networks show best results when the number of data points or images are very large in number. Loss values of the CNN model on the test dataset are represented in Fig \ref{loss}. Classification accuracy of individual fault parameters obtained from RF classifier are represented in Fig \ref{fault}. Classification accuracy for S.O. is the highest for RGB images while C.O. are better classified in grayscale images. Also, a very small degree of antenna tilt results in a noticeable change in the RSRP map images therefore enabling RF to better classify them as compared to TxP faults.

\section{Conclusion and Future Work}

Fault diagnosis is a highly desirable feature in emerging
complex and dense cellular networks to not only reduce
operational costs but also to improve quality of experience. In
this paper, we present a first of its kind neuromorphic based AI
model fault diagnosis solution that outperforms the prevailing,
CNN, RF and NB fault diagnosis schemes in existing literature.
The key advantage of proposed solution is that, it does not
require human experts for manual feature extraction unlike
existing solutions in literature. This advantage is achieved by
feeding RSRP data in form of images to train the models
instead of conventional approach of using raw fault data. In real
networks, these images can be created using MDT reports or
RSRP measurements gathered from other sources. The results
indicate that the proposed model can flag all seven faults in
the fed images. This framework can be used as a part of the
self-healing module in emerging networks. For future works,
we plan to investigate the performance of the proposed solution
in the presence of multiple faults concurrently as well as fault
detection.

\section*{Acknowledgment}
\vspace{-0.1cm}
This material is based upon work supported by the National Science Foundation under Grant Numbers 1559483, 1619346 and 1730650. For more details, please visit www.AI4Networks.com 
\vspace{-0.2cm}
\bibliographystyle{IEEEtran}
\bibliography{references_df}

\end{document}